\documentclass[aps,showpacs,preprint,floatfix]{revtex4}

\usepackage{amsfonts,amsmath,amssymb,mathrsfs,amsthm}
\usepackage{dsfont}
\usepackage{graphicx}
\usepackage{setspace}
\usepackage{url}
\usepackage{bm}
\usepackage{citesort,bm,amsfonts,amssymb}
\newcommand{\bd}[1]{\boldsymbol{#1}}
\usepackage{color}
\usepackage{ulem}

\begin{document}

\title{Duality of reduced density matrices and their eigenvalues}

\author{Christian Schilling}
\email{christian.schilling@physics.ox.ac.uk}
\affiliation{Clarendon Laboratory, University of Oxford, Parks Road, Oxford OX1 3PU, United Kingdom}

\author{Rolf Schilling}
\email{rschill@uni-mainz.de}
\affiliation{$^2$Institut f\"ur Physik, Johannes Gutenberg-Universit\"at Mainz, Staudinger Weg 9, D-55099 Mainz, Germany}

\date{\today}

\begin{abstract}
For states of quantum systems of $N$ particles with harmonic interactions we prove that each reduced density matrix $\rho$ obeys a duality condition. This condition implies duality relations for the eigenvalues $\lambda_k$ of $\rho$ and relates a harmonic model with length scales $\ell_1,\ell_2, \ldots, \ell_N $  with another one with inverse lengths $1/\ell_1, 1/\ell_2,\ldots, 1/\ell_N$. Entanglement entropies and correlation functions  inherit duality from $\rho$.  Self-duality can only occur for noninteracting particles in an isotropic harmonic trap.
\end{abstract}


\maketitle

\section{Introduction}

Duality is an important concept in different branches of physics. It relates physical quantities and physical behavior of two different systems with each other. Prominent examples are the particle-wave duality in quantum mechanics and the duality of Maxwell's electrostatic and magnetostatic theory obtained by a Lorentz transformation from a rest frame to a moving one. The particle-hole duality (see e.g.~\cite{1}) for models of electrons in solid state physics and the Kramers-Wannier duality \cite{2} in statistical physics are further examples. The latter relates the free  energy of an Ising model with nearest neighbor coupling $J$ on a $2$-dimensional lattice $\mathcal{L}$ to the free energy of an Ising model with coupling $J^*$ on the dual lattice  $\mathcal{L}^*$. In particular, it links the low temperature behavior of the lattice system $\mathcal{L}$ to the high temperature properties of the dual system. For a self-dual lattice, i.e.~$\mathcal{L} = \mathcal{L}^*$, like the square lattice, the duality allows to determine the critical point for the phase transition of the corresponding Ising model \cite{2}.

An interesting duality has been discovered recently \cite{4,4a} for the Moshinsky atom \cite{4b}, a two electron model,
in three dimensions described by the hamiltonian $\hat{H}=\frac{\hat{\vec{p}}_1^{\,2}}{2m}+\frac{\hat{\vec{p}}_2^{\,2}}{2m}+ \frac{1}{2}m\omega^2 (\hat{\vec{x}}_1^{\,2}+\hat{\vec{x}}_2^{\,2}\big) + \frac{1}{2} D\,(\hat{\vec{x}}_1- \hat{\vec{x}}_2)^2$. Here $m$ is the electron mass, $m\omega^2$ the force constant of an external harmonic potential and $D$ the coupling constant of the harmonic interactions.
By introducing the dimensionless coupling constant $\Lambda=D/(m\omega^2)$ a duality relation
\begin{equation}\label{eq1}
S_q(\Lambda) = S_q(\Lambda ')
\end{equation}
for the R$\acute{e}$nyi entropy $S_q(\Lambda)=(1-q)^{-1}\ln\big[�\mbox{Tr}[(\rho_{1}(\Lambda)^q]\big],  \ q \neq 1$, has been found \cite{4}
for the 1-particle reduced density operator ($1$-RDO) $ \rho_{1}(\Lambda)$ of the ground state.
Condition (\ref{eq1}) means that for any given $\Lambda \in (-\frac{1}{2},0)$  there exists a unique coupling constant $\Lambda '(\Lambda) \in (0,\infty)$ (see Ref.\cite{4}) such that $S_q(\Lambda) = S_q(\Lambda '(\Lambda))$.
For the ground state of a generalized\footnote{Since the Moshinsky atom is a system of $2$ electrons we call its generalization to $N$ electrons generalized Moshinsky atom.} Moshinsky atom of three electrons in one dimension a similar duality has been observed for the natural occupation numbers $\lambda _k^{(1)}$   \cite{5,8}
\begin{equation}\label{eq2}
\lambda _k^{(1)} (\ell _- /\ell _+) = \lambda _k ^{(1)} (\ell_{-}^{*} / \ell_{+}^{*}) ,\    k \in \mathbb{N}
\end{equation}
with $[\ell_{-}^{*} / \ell_{+}^{*}](\ell _- ,\ell _+)= \ell _+ /\ell _-$.
$\ell_+$ and $\ell _-$ are two natural length scales related to both coupling constants of the model.
 Numerical analysis \cite{6} has provided strong evidence that this kind of duality is not restricted to the ground state of the Moshinsky atoms with two or three electrons, but holds for any number $N$ of electrons, and for all its eigenstates.
In this communication we will answer three questions

\begin{enumerate}
\item What is the origin of that duality?
\item Does it hold for any harmonic model and for any $N$-particle eigenstate?
\item Is the duality also given for the eigenvalues of the $M$-particle reduced density operator ($M$-RDO) with $M>1$
      and what is the role of the exchange symmetry?
\end{enumerate}

These questions are of fundamental relevance since reduced density operators (RDO) and their eigenvalues play an important role in atomic physics and quantum chemistry \cite{9,10,11}. They also have attracted strong attention in quantum information theory where the so-called quantum marginal problem is studied (see  Refs.\cite{12,15,16,17,18,19,20}). It asks whether given density operators (marginals) for subsystems of a multipartite quantum system are compatible in the sense that they can arise from a common total state. One of the prime examples originating from quantum chemistry is the $M$-particle $N$-representability problem \cite{9,10}, the problem of describing the family of $M$-RDO which can arise from an antisymmetric $N$-particle state. Stimulated by the results of Borland and Dennis \cite{13} and Ruskai \cite{14}, Klyachko \cite{15,16} in a ground-breaking work solved recently the $1$-particle $N$-representability problem. His solution has revealed so-called generalized Pauli constraints, restrictions on the eigenvalues of the $1$-RDO of fermionic systems, which are significantly strengthening Pauli's famous exclusion principle. Accordingly, answers to the questions raised above will add a new facet to the properties of RDO and their eigenvalues and therefore to this field of research.

\section{Harmonic model}
We consider a model for  particles with mass $m$ interacting harmonically. The corresponding hamiltonian is given by
\begin{equation}\label{eq3}
\hat{H} = \frac {1} {2m} \sum \limits _{i=1} ^N \hat{p} _i ^2 + \frac 1 2 \sum \limits _{1 \leq i , j \leq N}  D_{ij} \hat{x}_i \hat{x}_j \, .
\end{equation}
$\hat{x}_i$ and $\hat{p}_i$  are the position and momentum operators, respectively. $N$ is the number of degrees of freedom.
The interaction matrix $\textbf{D} = (D_{ij})$ is assumed to be positive definite and real-symmetric.
Note that this hamiltonian holds in any dimension. For $d>1$ the components  $\hat{x}_{r,\alpha}$ and $\hat{p}_{r,\alpha}$, for $r=1,\ldots,N'$ and $\alpha =1,\ldots,d$ of the position and momentum operators, respectively, should be relabeled, the same should be done for the  matrix, $ D_{ij}=D_{r\alpha,s\beta}$, and $N$ in Eq.~(\ref{eq3}) is given by $N=d \cdot N'$. Of course,  possible translational and  rotational symmetry would impose additional constraints on $\textbf{D}$.

For the case of identical particles the potential energy, the second term in Eq.~(\ref{eq3}), is invariant under particle permutations.
This restricts $\{D_{r\alpha,s\beta}\}$  to two independent matrices
$\textbf{D}^{(1)}$ and $\textbf{D}^{(2)}$ and the potential energy takes the form $\frac{1}{2}\sum_{r=1}^{N'}\,\hat{\vec{x}}_r^{\,t} \textbf{D}^{(1)}\, \hat{\vec{x}}_r + \frac{1}{2}\sum_{1\leq r <s\leq N'}(\hat{\vec{x}}_r-\hat{\vec{x}}_s)^t \textbf{D}^{(2)} (\hat{\vec{x}}_r-\hat{\vec{x}}_s)$, where
$D^{(2)}_{\alpha,\beta} = -D_{1 \alpha,2 \beta}$ and $D^{(1)}_{\alpha,\beta} \equiv D_{1\alpha,1\beta}-(N'-1)D^{(2)}_{\alpha,\beta}$ and
$\hat{\vec{x}}_r\equiv(\hat{\vec{x}}_{r 1},\ldots,\hat{\vec{x}}_{r d})^t$.
Consequently,  model (\ref{eq3}) (with corresponding $\textbf{D} > 0$) can also arise for translationally invariant models (i.e.~$\textbf{D}^{(1)}=\textbf{0}$) after a separation of the relative degrees of freedom from the free center of mass motion. Prominent examples for such harmonic systems are the Moshinsky atom \cite{4b} and its generalization to an arbitrary number of fermions \cite{7}, which were already mentioned in the introduction.
Note, hamiltonian (\ref{eq3})  can also be used to describe a $d$-dimensional  harmonic lattice  with $N'$ sites,  $r= 1,...,N' $  and dynamical matrix $\{D_{r\alpha,s\beta}\}$.

For the following technical analysis, to avoid annoying issues related to the dimensionality of physical quantities, we set several physical quantities as e.g.~$\hbar$ and $m$ to $1$ and treat any variable as e.g.~positions $x_i$, momenta $p_i$ and couplings $D_{ij}$ as dimensionless real numbers.
This could also be achieved by introducing an arbitrary length scale $L$ and dimensionless position operators $\hat{x}_i/L$, momentum operators
$\hat{p}_i/(\hbar/L)$ and hamiltonian $\hat{H}/(\hbar \Omega)$ with $\Omega\equiv\hbar/(L^2 m)$.
In position space $\hat{H}$ then takes the form
\begin{equation}\label{eq4}
H^{(x)} = - \frac {1}{2} \Delta_{\vec{x}} + \frac{1}{2} \vec{x}^t \textbf{D} \vec{x}\,
\end{equation}
where $\vec{x}\equiv (x_1,\ldots,x_N)^t \in \mathbb{R}^N$ and $\Delta_{\vec{x}}$ is the corresponding Laplace operator.
$H^{(x)}$ is densely defined on the separable Hilbert space $L^2(\mathbb{R}^N)$ of square-integrable functions $\Psi: \mathbb{R}^N\rightarrow\mathbb{C}$.
The corresponding time-independent Schr\"odinger equation reads
\begin{equation}\label{eq5}
H^{(x)} \Psi  = E \Psi \, .
\end{equation} \,
To determine the eigenenergies $E$ we just need to diagonalize the symmetric coupling matrix $\textbf{D}$,
\begin{equation}\label{eq8}
{\textbf{RDR}^t = \textbf{d}} \,,
\end{equation}
where $\textbf{d}\equiv \mbox{diag}(d_1,\ldots,d_N)$ is a diagonal matrix and $\textbf{R}$ an orthogonal matrix. The coordinate transformation from $\vec{x}$ to the pseudo-positions $ {\vec{y}}  = \textbf{R}\vec{x}$ is represented as a unitary transformation $U(\textbf{R})$ on $L^2(\mathbb{R}^N)$,
\begin{equation}\label{unitaryR}
(U(\textbf{R})\Psi\big)(\vec{y}) \equiv \Psi(\textbf{R}^{-1}\vec{y})\qquad,\, \forall \vec{y}\in \mathbb{R}^N\,.
\end{equation}
The transformed hamiltonian $H^{(y)}\equiv  U(\textbf{R})H^{(x)}U(\textbf{R})^\dagger$ takes the diagonal form
\begin{equation}\label{eq11}
H^{(y)} = \sum \limits ^N _{\mu = 1} \frac{1}{2} \frac{1}{\ell_\mu} \,[- \frac {\partial ^2}{\partial(y_\mu/\ell_\mu)^2} + (y_\mu / \ell_\mu)^2] \, .
\end{equation}
where we introduced the dimensionless length scales $\ell_\mu = \frac{1}{\sqrt{d_\mu}}$.
The normalized eigensolutions of $H^{(y)}$ are products of $1$-particle states $\varphi_\ell^{(\nu)}(y)$
\begin{equation}\label{eq12}
\Phi_{\bd{\ell}}^{(\boldsymbol{\nu})}(\vec{y}) = \prod \limits^N _{\mu = 1} \varphi ^{(\nu_\mu)}_{\ell_\mu}(y_\mu)
\end{equation}
with $\boldsymbol{\nu} = (\nu_1,\ldots,\nu_N), \; \nu_\mu \in \mathbb{N}_0$, $\bd{\ell}\equiv(\ell_1,\ldots,\ell_N)$ and
\begin{equation}\label{eq13}
\varphi _\ell^{(\nu)} (y) = \pi ^{- \frac 1 4} \ell ^{- \frac 1 2} (2^\nu \nu ! )^{-\frac 1 2} H_\nu (\frac y \ell) \exp{\big(-\frac {y^2}{2 \ell^2}\big)}\,.
\end{equation}
$H_\nu(z)$ are the Hermite polynomials. Then, the eigenstates of $H^{(x)}$ follow as
\begin{equation}\label{eq14}
\Psi_{\bd{\ell}}^{(\bd{\nu})}(\vec{x})= \big(U(\textbf{R})^\dagger\Phi_{\bd{\ell}} ^{(\bd{\nu})}\big)(\vec{x}) = \Phi_{\bd{\ell}} ^{(\bd{\nu})}(\textbf{R}\vec{x})\qquad,\,\forall \vec{x}\in \mathbb{R}^N.
\end{equation}
For later purpose, we have made explicit the dependence of $\Psi_{\bd \ell}^{(\bd{\nu})}$ on $\bd{\ell}\equiv (\ell_1,\ldots,\ell_N)$.
Moreover, from Eq.~(\ref{eq12}) and the explicit form of the $1$-particle states (\ref{eq13}) we can infer a homogeneous structure,
\begin{equation}\label{homog}
\Psi_{\bd{\ell}}^{(\bd{\nu})}(\vec{x}) = \alpha^{-\frac{N}{2}}\Psi_{\frac{\bd{\ell}}{\alpha}}^{(\bd{\nu})}\big(\frac{\vec{x}}{\alpha}\big) \equiv \big(V_N(\alpha)\Psi_{\frac{\bd{\ell}}{\alpha}}^{\bd{(\nu)}}\big)(\vec{x})\qquad,\, \forall \alpha \in \mathbb{R}^{+}\,.
\end{equation}
Here we have introduced the unitary rescaling operator $V_N(\alpha)$, which is rescaling the length by a factor $\alpha$ and is local,
$V_N(\alpha) = V_1(\alpha)^{\otimes^N}$, where $V_1(\alpha)$ the rescaling operator on $L^2(\mathbb{R})$.

Due to the harmonic character of the hamiltonian (\ref{eq4}) it is worth studying the Schr\"odinger equation (\ref{eq5}) also in the Fourier space.
For this we introduce the Fourier transformation, a linear bounded operator on the space of Schwartz functions $\varphi:\mathbb{R}\rightarrow \mathbb{C}$,
\begin{equation}\label{fourier}
\big(\mathcal{F}_1\varphi\big)(p)  \equiv \!\int_{\mathbb{R}}\mathrm{d}x\,\varphi(x)\,e^{-i \,x\,p}\,.
\end{equation}
For all $k \in \mathbb{N}$, $\mathcal{F}_1$ gives rise to Fourier operator $\mathcal{F}_k$ acting on Schwartz functions $\Phi:\mathbb{R}^k\rightarrow \mathbb{C}$ which is extended uniquely to the space $L^2(\mathbb{R}^k)$ according to the Plancherel theorem \cite{21}.
By identifying $L^2(\mathbb{R}^k)\cong \big(L^2(\mathbb{R})\big)^{\otimes^k}$ the local structure $\mathcal{F}_k = \mathcal{F}_1^{\otimes^k}$ w.r.t~ the $1$-particle Hilbert spaces $L^2(\mathbb{R})$ is obvious.

Applying $\mathcal{F}_N$ to Eq.~(\ref{eq5}) leads to the Schr\"odinger equation in momentum space
\begin{equation}\label{Hp}
H^{(p)} \tilde{\Psi} = E \tilde{\Psi}\,
\end{equation}
with $\tilde{\Psi}\equiv \mathcal{F}_N \Psi$, the Fourier transformation of $\Psi$ and
\begin{equation}\label{eq6}
H^{(p)} \equiv \mathcal{F}_N H^{(x)} \mathcal{F}_N^{\dagger}= - \frac{1}{2} \nabla_{\vec{p}}^t\,\textbf{D}\,\nabla_{\vec{p}} + \frac{1}{2}\,\vec{p\,}^2\, .
\end{equation}
Due to the unitarity of $\mathcal{F}_N$, $H^{(x)}$ and $H^{(p)}$ have the same spectrum. Below we will see that this dual description, using either the position or the momentum space, is the origin of the duality of the $M$-RDO and of its eigenvalues.

The eigenvalue problem for $H^{(p)}$ can be solved similarly as that for $H^{(x)}$. With the pseudo-momenta $\vec{\pi}= \textbf{R}\vec{p}$ and Eqs.~(\ref{eq8}), (\ref{unitaryR}), $H^{(p)}$ is transformed to
\begin{equation}\label{eq17}
H^{(\pi)} \equiv U(\textbf{R}) H^{(p)} U(\textbf{R})^\dagger = \sum \limits ^N _{\mu =1} \frac{1}{2} \frac{1}{\tilde{\ell}_{\mu}} [- \frac {\partial ^2} {\partial(\pi _\mu /\tilde{\ell}_\mu)^2} + (\pi_\mu/\tilde{\ell}_\mu)^2]\, ,
\end{equation}
with the reciprocal (dimensionless) length scales
\begin{equation}\label{lengthduality}
\tilde{\ell}_{\mu} \equiv \frac{1}{\ell_\mu}\qquad, \forall \mu =1,\ldots,N\,.
\end{equation}
Notice that the hamiltonians (\ref{eq11}) and (\ref{eq17}) are identical up to a swapping of $\bd{\ell}$ and $\tilde{\bd{\ell}}$. Since $H^{(x)}= U(\textbf{R})^\dagger H^{(y)}  U(\textbf{R})$ and $H^{(p)}= U(\textbf{R})^\dagger H^{(\pi)}  U(\textbf{R})$ the same also holds for $H^{(x)}$ and $H^{(p)}$. Consequently, we find
\begin{equation}\label{Hpeigen}
H^{(p)} \Psi_{\tilde{\bd \ell}}^{(\bd{\nu})} = E_{\tilde{\bd\ell}}^{(\bd{\nu})} \Psi_{\tilde{\bd \ell}}^{(\bd{\nu})}\,.
\end{equation}
Relating this to Eq.~(\ref{Hp}) with $\tilde{\Psi}\rightarrow \tilde{\Psi}_{\bd{\ell}}^{(\bd{\nu})}$ finally yields the important relation
\begin{equation}\label{eq19b}
\mathcal{F}_N \Psi_{\bd{\ell}}^{(\bd{\nu})} = \Psi_{\tilde{\bd{\ell}}}^{(\bd{\nu})} \,.
\end{equation}
It states that applying the Fourier operator $\mathcal{F}_N$ to an eigenstate of the hamiltonian (\ref{eq4}) does only lead to a rescaling of the length scales, i.e.~a replacement of $\bd{\ell}$ by $\tilde{\bd{\ell}}$ according to Eq.~(\ref{lengthduality}).

\section{Reduced density operators}
Instead of elaborating on eigenfunctions we focus on reduced density operators. To keep the notation simple we restrict ourselves for most of this section to $N$-particle systems with fermionic or bosonic exchange symmetry. The generalization to arbitrary/no symmetries will be discussed at the end of this section.

First, for a state $\Psi \in L^2(\mathbb{R}^N)$ we define the corresponding density operator $\rho:L^2(\mathbb{R}^N)\rightarrow L^2(\mathbb{R}^N)$,
\begin{equation}\label{NRDO}
\rho\,\Phi = \langle \Psi,\Phi\rangle_N\,\Psi\qquad,\,\forall \Phi \in L^2(\mathbb{R}^N).
\end{equation}
Here, $\langle \cdot, \cdot\rangle_N$ is the inner product on $L^2(\mathbb{R}^N)$ . By making use of the tensor product structure $L^2(\mathbb{R}^N)\cong L^2(\mathbb{R})^{\otimes^N}$ we can define for $m=1,\ldots,N-1$ the $M$-particle reduced density operator ($M$-RDO) $\rho^{(M)}$ of $\rho$,
\begin{equation}\label{mRDO}
\rho^{(M)} \equiv \mbox{Tr}_{N-M}[\rho]\,,
\end{equation}
where we introduced the partial trace $\mbox{Tr}_{N-M}[\cdot]$ (see e.g.~\cite{23}). Due to the exchange symmetry $\mbox{Tr}_{N-M}[\rho]$ is independent of the choice of the $N-M$ particles that are traced out. Since $\rho^{(M)}:L^2(\mathbb{R}^M)\rightarrow L^2(\mathbb{R}^M)$ we can study its eigenvalue equation on $L^2(\mathbb{R}^M)$,
\begin{equation}\label{eigeneqk}
\rho^{(M)} \chi^{(M)} = \lambda^{(M)} \chi^{(M)}\,.
\end{equation}
In quantum chemistry \cite{9} $\rho^{(M)}$ is typically represented w.r.t.~a basis of $L^2(\mathbb{R}^M)$ which leads to an infinite-dimensional $M$-particle reduced density \textit{matrix}.

The $1$-RDO, $\rho^{(1)}$, plays a particular role, since its eigenfunctions $\{\chi_k^{(1)}\}$ provide a $1$-particle description of the $N$-particle state $\Psi^{(N)}$ with the corresponding eigenvalues $\{\lambda _k^{(1)}\}$ as occupation numbers, where $ k \in \mathbb{N}$. In case of a $2$-particle observable $\hat{A}^{(2)}$ for a system of $N$ identical particles, $\rho^{(2)}$ arising from $\Psi^{(N)}$ is sufficient in order to calculate the expectation value $\langle\hat{A}^{(2)}\rangle_{\Psi^{(N)}}$. One important concept based on that is the simplified calculation of fermionic ground state energies: For hamiltonians with  $2$-particle interactions, the ground state can be obtained by minimizing the energy expectation value just over $2$-RDO $\rho^{(2)}$, which arise from  $N$-fermion states $\Psi^{(N)}$. The underlying set of states $\rho^{(2)}$ is much smaller than that of $N$-fermion quantum states, but the problem is then to determine the set of possible $\rho^{(2)}$'s, which is known as the $2$-particle $N$-representability problem \cite{10}. This makes obvious why reduced density operators particular for $M=1$ and $M=2$ are intensively studied in atomic physics and quantum chemistry, as well as, in quantum information theory as explained in the last paragraph of Section 1.

Before applying the concept of RDO to the eigenstates of our harmonic model (\ref{eq3}) we comment on the relation of Fourier operators $\mathcal{F}_1^{\otimes^N}$ and partial traces. Since $\mathcal{F}_k \equiv \mathcal{F}_1^{\otimes^k}$ is a unitary operator
and partial traces are respecting the local tensor product structure of $L^2(\mathbb{R}^N) \cong L^2(\mathbb{R})^{\otimes^N}$
we have
\begin{equation}\label{mRDOFourier}
\mbox{Tr}_{N-M}[\mathcal{F}_1^{\otimes^N}\,\rho \,(\mathcal{F}_1^\dagger)^{\otimes^N}]
= \mathcal{F}_1^{\otimes^M}\mbox{Tr}_{N-M}[\rho] (\mathcal{F}_1^\dagger)^{\otimes^M}
=  \mathcal{F}_k  \rho^{(M)} \mathcal{F}_M^\dagger\,.
\end{equation}
This means that if $N$-particle density operators $\rho$ and $\tilde{\rho}$ are conjugate by $\mathcal{F}_N$, $\tilde{\rho}=\mathcal{F}_N\,\rho \,\mathcal{F}_N^\dagger$, their $M$-RDO $\tilde{\rho }^{(M)}$ and $\rho^{(M)}$ are conjugate by $\mathcal{F}_M$, $\tilde{\rho}^{(M)} = \mathcal{F}_M \rho^{(M)}  \mathcal{F}_M^\dagger$.

Let us return to the harmonic model. For any arbitrary eigenstate $\Psi_{\bd{\ell}}^{(\bd{\nu})}$ we define its corresponding density operator
$\rho_{\bd{\ell}}$ and its $M$-RDO $\rho_{\bd{\ell}}^{(M)}$ according to Eqs.~(\ref{NRDO}) and (\ref{mRDO}). The same should  also be done for the Fourier transformation $\mathcal{F}_N \Psi_{\bd{\ell}}^{(\bd{\nu})}$ of $\Psi_{\bd{\ell}}^{(\bd{\nu})}$ which yields $\tilde{\rho}_{\bd{\ell}}$ and its $M$-RDO $\tilde{\rho}_{\bd{\ell}}^{(M)}$. Note that we skip the superscript $\bd{\nu}$ to keep our notation simpler.

Since the Fourier transformation $\mathcal{F}_k$ is not only a unitary operator but even an isometry (Plancherel theorem \cite{21}) we find for any $\Phi \in L^2(\mathbb{R}^N)$ by recalling Eq.~(\ref{NRDO}) and $\tilde{\Psi}\equiv \mathcal{F}_N \Psi$
\begin{equation}
\tilde{\rho}_{\bd{\ell}}\,\Phi \equiv \langle \tilde{\Psi}_{\bd{\ell}}^{(\bd{\nu})} , \Phi\rangle_N \tilde{\Psi}_{\bd{\ell}}^{(\bd{\nu})}
= \langle \Psi_{\bd{\ell}}^{(\bd{\nu})} , \mathcal{F}_N^{-1}\Phi\rangle_N \tilde{\Psi}_{\bd{\ell}}^{(\bd{\nu})}
= \mathcal{F}_N  \rho_{\bd{\ell}} \mathcal{F}_N^{-1}\,\Phi \,.
\end{equation}
Hence, we have $\tilde{\rho}_{\bd{\ell}} = \mathcal{F}_N  \rho_{\bd{\ell}} \mathcal{F}_N^\dagger$ and according to the comment in the first two lines below Eq.~(\ref{mRDOFourier})
\begin{equation}\label{mRDOconj}
\tilde{\rho}_{\bd{\ell}}^{(M)} = \mathcal{F}_M  \rho_{\bd{\ell}}^{(M)}\, \mathcal{F}_M^\dagger\,.
\end{equation}
Eq.~(\ref{mRDOconj}) together with Eq.~(\ref{eq19b}) implies for the $M$-RDO of any eigenstate of hamiltonian (\ref{eq4})
\begin{equation}\label{eq22a}
\rho_{\tilde{\bd{\ell}}}^{(M)} = \mathcal{F}_M  \rho_{\bd{\ell}}^{(M)}\, \mathcal{F}_M^\dagger\,.
\end{equation}

This result for $N$-particle systems with fermionic or bosonic exchange symmetry can be generalized to
arbitrary $N$-particle states. The only subtle difference is that without exchange symmetry the $M$-RDO depends
on the choice of the $(N-M)$ particles which are integrated out in Eq.~(\ref{mRDO}). To take this into account, we obtain now
for every choice of particles $\textbf{m}\equiv(m_1,\ldots,m_M)$ a corresponding $M$-RDO $\rho_{\bd{\ell}}^{(\textbf{m})}$.
For each of those $\Big(\!\begin{array}{c}N\\M\end{array}\!\Big)$ $M$-RDO the result Eq.~(\ref{eq22a}) holds separately,
\begin{equation}\label{eq22b}
\rho_{\tilde{\bd{\ell}}}^{(\textbf{m})} = \mathcal{F}_M  \rho_{\bd{\ell}}^{(\textbf{m})}\, \mathcal{F}_M^\dagger\,.
\end{equation}
Equation (\ref{eq22b}) is the duality relation for the $M$-RDO corresponding to a general $N$-particle state of a system described by hamiltonian (\ref{eq4}).

We close this section by deriving a homogeneous structure for the $M$-RDO following from the homogeneity relation
(\ref{homog}) for the $N$-particle eigenstates $\Psi_{\bd{\ell}}^{(\bd{\nu})}$. Recalling the unitary rescaling
operator $V_N(\alpha)$ from Eq.~(\ref{homog}), Eq.~(\ref{NRDO}) implies
\begin{equation}
\rho_{\bd{\ell}}\Phi =  \langle V_N(\alpha)^\dagger \Psi_{\bd{\ell}} , V_N(\alpha)^\dagger \Phi\rangle_N \Psi_{\bd{\ell}}
=  \langle \Psi_{\frac{\bd{\ell}}{\alpha}} , V_N(\alpha)^\dagger \Phi\rangle_N V_N(\alpha) \Psi_{\frac{\bd{\ell}}{\alpha}} \,.
\end{equation}
This yields the homogeneous structure
\begin{equation}\label{homogNRDO}
\rho_{\bd{\ell}} =  V_N(\alpha) \rho_{\frac{\bd{\ell}}{\alpha}} V_N(\alpha)^\dagger\qquad,\,\forall \alpha \in \mathbb{R}^+\,.
\end{equation}
By using the local structure $V_N(\alpha) = V_1(\alpha)^{\otimes^N}$, Eq.~(\ref{mRDO}), the unitarity of $V_1(\alpha)$ and
the properties of the partial trace it is an elementary exercise to show that the $M$-RDO inherit the same homogeneous structure (\ref{homogNRDO}),
\begin{equation}\label{homogMRDO}
\rho_{\bd{\ell}}^{(\textbf{m})} =  V_M(\alpha) \rho_{\frac{\bd{\ell}}{\alpha}}^{(\textbf{m})} V_M(\alpha)^\dagger\qquad,\,\forall \alpha \in \mathbb{R}^+\,,
\end{equation}
where $V_M(\alpha) = V_1(\alpha)^{\otimes^M}$ and $M= |\textbf{m}|$.

\section{Duality for eigenvalues and interaction matrices}
In this section we work out the consequences of the duality condition (\ref{eq22b}) of $M$-RDO.

Since the Fourier operator $\mathcal{F}_k$ is unitary the spectral consequences of result (\ref{eq22b}) are obvious:
Both density operators $\rho_{\bd{\ell}}^{(\textbf{m})}$ and $\rho_{\tilde{\bd{\ell}}}^{(\textbf{m})}$ have the same eigenvalues.
Moreover, due to the homogeneous structure (\ref{homogMRDO}) they depend only on the projective point $[\bd{\ell}]$ of $\bd{\ell}$ which is the equivalence class of $\bd{\ell}$ w.r.t.~the relation $\bd{\ell}\sim \bd{\ell}' :\Leftrightarrow \bd{\ell} = \alpha \bd{\ell}'$, for some $\alpha \in \mathbb{R}$.
We denote these eigenvalues by $\lambda_k^{(\textbf{m})}([\bd{\ell}])$, $k\in \mathbb{N}$, and order them decreasingly, $\lambda_k^{(\textbf{m})}([\bd{\ell}])\geq \lambda_{k+1}^{(\textbf{m})}([\bd{\ell}])$. Then, we have
\begin{equation}\label{NONduality}
\lambda_k^{(\textbf{m})}([\bd{\ell}]) = \lambda_k^{(\textbf{m})}([\tilde{\bd{\ell}}])\qquad,\,\forall k \in \mathbb{N}\,,
\end{equation}
where $\tilde{\bd{\ell}}\equiv \tilde{\bd{\ell}}(\bd{\ell})$ is given by Eq.~(\ref{lengthduality}) and we recall that we have skipped the superscript
$\bd{\nu}$ in (\ref{NONduality}), which labels the underlying $N$-particle energy eigenstate.

Identity (\ref{NONduality}) is our main result and it is the generalization of the duality condition (\ref{eq2}) found in Ref.\cite{5}. It does not depend on the use of dimensionless quantities and is valid for \textit{arbitrary} number $N$ of particles, for \textit{any} eigenstate state of the harmonic model (3) and for any choice of the subsystem $\textbf{m}\equiv(m_1,\ldots,m_M) \subset \{1,\ldots,N\}$ of particles $m_1,\ldots,m_M$. Since the $d$-dimensional harmonic model can also be represented by hamiltonian (3), the duality relation also holds for \textit{arbitrary} dimensions.

Since \{$\ell_{\mu}/\ell_{\mu+1}\}_{\mu=1,...,N-1}$ is in a  one-to-one relation  with the projective point $[\bd{\ell}]$,
we can follow Ref.\cite{5} and  introduce
$\delta _\mu =  \ln (\ell_\mu/\ell_{\mu+1}) \; , \mu =1,\ldots,N-1$. Then
Eq.~(\ref{NONduality}) becomes
\begin{equation}\label{eq26}
\lambda_k^{(\textbf{m})} (\delta_1,\ldots,\delta_{N-1}) \equiv \lambda_k^{(\textbf{m})}(-\delta_1,\ldots,-\delta_{N-1}),\ k \in \mathbb{N}
\end{equation}
where we used for the function on $\{\delta_\mu\}$ the same symbol $\lambda_k^{(\textbf{m})}$ as for the function on $[\bd{\ell}]$ in order to avoid further overloading  notation. The duality in form of Eq.~(\ref{eq26}) implies that the eigenvalues $\lambda_k^{(\textbf{m})}$ are even functions in $\{\delta_\mu\}$. Since $\delta_\mu =0$ for $\ell_\mu \equiv \ell_{\mu +1}$, which corresponds to vanishing interactions, the dimensionless parameters $\delta_\mu$ are a measure of the interaction. If the decreasingly ordered eigenvalue functions $\lambda_k(\delta_1,\ldots,\delta_{N-1})$ are also analytic at $\delta_1=0,\ldots ,\delta_{N-1}=0$ their Taylor series around that point contain only even order terms in each $\delta_\mu$ \footnote{Although the spectrum of any RDO $\rho_{\textbf{m}}$
is analytic in $\{\delta_\mu\}$ the ordering $\lambda_k \geq \lambda_{k+1}$ of the eigenvalues leads to non-analytic behavior at $\delta_1,\ldots,\delta_{N-1}=0$ whenever two hypersurfaces $\lambda_k(\delta_1,\ldots,\delta_{N-1})$ and $\lambda_{k+1}(\delta_1,\ldots,\delta_{N-1})$ do cross at that point.}.
This has been the case for the ground state of the generalized Moshinsky atom investigated  in Ref.~\cite{5} and has significantly simplified the perturbational analysis of the natural occupation numbers for weak couplings.

In the case of a $r$-fold degeneracy of the length scales $\ell_\mu$, i.e.~there exits $\mu$ such that $\ell_\mu = \ell_{\mu+1} = \cdots = \ell_{\mu +r-1}$,  it is $\ell_{\mu +\nu}/\ell_{\mu + \nu+1} = 1$ for $\nu = 0,1,\ldots,r-2$. Then, the number of variables in
(\ref{eq26}) is effectively reduced by $(r - 1)$. This holds for the harmonic model studied in Ref.~\cite{5} where the degeneracy has led to $\ell_1 \equiv \ell_{-}$ and $\ell_2 = \ell_3 = \cdots = \ell_N \equiv \ell_{+}$.  Consequently, duality condition (\ref{NONduality}) for fermionic or bosonic states and $M=1$ reduces to Eq.~(\ref{eq2}).

If hamiltonian (\ref{eq3}) arises as a \textit{periodic} harmonic lattice model with long range interactions nothing changes qualitatively, if $D_{r\alpha,s\beta} \to 0$ with the distance between sites $r$ and $s$ going to infinity. However, if  $D_{r\alpha,s\beta}=const.$  for all $r \neq s$ then, depending on the lattice symmetry and dimension $d$ there exist $\mu$ such that $\ell_{\mu}/\ell_{\mu+1}=1$ ,
leading to a reduction of the number of independent variables of $\lambda_k^{(\textbf{m})}$ .

\section{Discussion and conclusions}
We have proven analytically that the recently discovered duality of the R$\acute{e}$nyi entropy (Ref.~\cite{4}) and of the natural occupation numbers (Ref.~\cite{5}) of the ground state of Moshinsky-type atoms with two and three electrons, respectively, is \textit{generic} for arbitrary $N$-particle systems with harmonic interactions, in any dimension. It originates from the duality of wave mechanics in position and momentum space in combination with the harmonic interactions.
The dualities (\ref{eq22b}) and (\ref{NONduality}) are valid for all RDO of any arbitrary \textit{pure} $N$-particle eigenstate. A possible fermionic or bosonic exchange symmetry for the case of identical particles is irrelevant.
Moreover, duality also holds for arbitrary $N$-particle pure states of system (\ref{eq3}), i.e.~states of the form $\Psi_{\bd{\ell}} =\sum_{\bd{\nu}}\,c_{\bd{\nu}} \Psi^{(\bd{\nu})}_{\bd{\ell}}$, which even generalizes to mixed states. As a consequence, for spinful particles described by hamiltonian (\ref{eq3}), the duality (\ref{NONduality}) holds for the eigenvalues of any orbital RDO. Moreover, it also holds for any full (i.e.~spin including) RDO.

What is the physical implication of that duality?

The kind of duality found here is quite analogous to the Kramers-Wannier duality, because the duality relation (\ref{eq22b}) or (\ref{NONduality}) allows to relate a harmonic model (\ref{eq3}) with interaction matrix  $\textbf{D}$ with a \textit{dual} model with interaction  matrix ${\textbf{D}}^{*}$, i.e.~a model with \textit{different} coupling constants. From section $2$ it is clear how the dual model looks like. First notice that the  hamiltonian
\begin{equation}
\hat{H}_{\alpha,\beta}(\textbf{D}) = \alpha \hat{T} + \beta \frac{1}{2}\,\hat{\vec{x}}^t\,\textbf{D}\,\hat{\vec{x}}\,,
\end{equation}
with $\hat{T}$ the kinetic energy operator, $\alpha,\beta \in \mathbb{R}^+$ coincides with that in Eq.~(\ref{eq3}),
$\hat{H}_{1,1}(\textbf{D})$, up to a rescaling of the energy and length scale. Therefore, we call $\hat{H}_{\alpha,\beta}(\textbf{D})$ and $\hat{H}_{\alpha',\beta'}(\textbf{D})$  equivalent. The spectrum of any RDO $\rho^{(\textbf{m})}$ of the $\boldsymbol{\nu}$-th eigenstate of $\hat{H}_{\alpha,\beta}$ is independent of $\alpha$ and $\beta$ and depends only on the equivalence class $[\hat{H}(\textbf{D})]$. Due to this ambiguity the dual model $\hat{H}(\textbf{D}^*)$ is only uniquely defined on the level of those equivalence classes. Recalling Eqs.~(\ref{eq8}),(\ref{lengthduality}) and $l_\mu \equiv \frac{1}{\sqrt{d_\mu}}$ we find
\begin{equation}\label{dualD}
\textbf{D}^*(\textbf{D}) = \textbf{D}^{-1} \,.
\end{equation}
A model $\hat{H}({\textbf{D}})$ is self-dual if its dual model belongs to the same equivalence class, i.e. $\hat{H}({\textbf{D}}^{*}({\textbf{D}})) \sim \hat{H}({\textbf{D}})$. In that case it is straightforward to show that Eq.~(\ref{dualD}) implies that $D_{mm} \equiv const$  and
$D_{mn} = 0$ for all $m \neq n$, which corresponds to noninteracting particles in an isotropic harmonic trap.
Summarizing, the duality (\ref{NONduality}) relates  two physical harmonic models corresponding to two equivalence classes of hamiltonians, $[\hat{H}(\textbf{D})]$ and $[\hat{H}(\textbf{D}^*(\textbf{D}))]$,
\begin{equation}
[\hat{H}(\textbf{D})] \leftrightarrow [\hat{H}(\textbf{D}^*(\textbf{D}))]\,,
\end{equation}
with $\textbf{D}^*(\textbf{D})$ given by Eq.~(\ref{dualD}).

As an example, consider identical particles in one dimension. As explained in section 2 there are only two coupling parameters $D^{(1)}$ and $D^{(2)}$.
Like in Refs.~\cite{5,8} there are just two length scales and their ratio depends on $D^{(2)}/D^{(1)}$, only. Then Eq.~(\ref{dualD}) leads to
%\begin{equation}\label{eq28}
%D_{0}^{*}/D^{*} = -  \frac 1 2 [(1- \frac 2 N)+  \frac 1 2 D_{0}/D] .
%\end{equation}
\begin{equation}\label{eq28}
{D^{(2)}}^{*}/{D^{(1)}}^{*} = -\frac{{D^{(2)}}/{D^{(1)}}}{1+N{D^{(2)}}/{D^{(1)}}}.
\end{equation}
Since the eigenvalues $D^{(1)}$ and $D^{(1)}+ND^{(2)}$ (with multiplicity $N-1$) of the interaction matrix $\textbf{D}$ are both positive
Eq.~(\ref{eq28}) relates two relative interaction strengths ${D^{(2)}}^{*}/{D^{(1)}}^{*}$ and ${D^{(2)}}/{D^{(1)}}$ with opposite sign.
If the original model with $\textbf{D}$ describes attractive particles the dual one with $\textbf{D}^*$ describes repulsive particles, and vice versa.

Although the original hamiltonian $\hat{H}({\textbf{D}})$ and its duals  $\hat{H}^{*}\equiv\hat{H}({\textbf{D}^{*}})$ have different spectra and different eigenfunctions, the eigenvalues of $\rho^{(\textbf{m})}$ of corresponding  quantum states of $\hat{H}$ and  $\hat{H}^{*}$, e.g.~their ground states,   are identical.  The same is true for any function of these eigenvalues, like  entanglement entropies and correlation functions. For the  specific models investigated in  Refs.~\cite{4,5}, the duality condition implies that the interactions in the dual  model are \textit{repulsive} (\textit{attractive}), if those of the original one  are \textit{attractive} (\textit{repulsive}) (see also Eq.~(\ref{eq28})). For the general model (\ref{eq3}) the interactions can be a mixture of attractive and repulsive terms under the constraint of positive definiteness of ${\textbf{D}}$.
Duality   can  interchange attractive and repulsive terms.
Despite the different nature of their interactions  this duality relates a generalized Moshinsky atom or
a harmonic lattice (not necessarily periodic)  with a \textit{dual} one  leaving the physical properties as entanglement, correlation functions, etc.~unchanged.
Whereas the Kramers-Wannier duality \cite{2} links low and high temperature behaviour,  the duality found in the present contribution connects strongly localized states with $\ell_\mu$ ``small'' with weakly localized states of the dual model for which $\ell ^{*}_\mu$ is ``large'', or vice versa.

%Since for an atom or a molecule with arbitrary pair interactions may be approximated by a harmonic model the present %duality, at least on a qualitative level may also allow to find "dual atoms" and "dual molecules". \\

%To conclude, we have analytically proven that the duality condition for the occupation numbers found recently, is a %general property of an arbitrary particle system with harmonic interactions. It is valid for the eigenvalues of the
%$M$-particle reduced density operator of any pure N-fermion/boson eigenstates $\Psi^N$, for all $M=1,2,\ldots,N$. Even %for the case without any specific symmetry under particle exchange our results still hold. There, the only subtle difference %is that there is no unique $M$-RDO since it matters now which $N-M$ particles are traced out. However, for \emph{each} %such $M$-RDO the duality holds. Moreover, if the dynamical matrix $\textbf{D}$ in Eq.~(\ref{eq3}) has a degenerate %spectrum also mixed energy eigenstates are possible. Also for their $M$-RDO the duality is still given. Clearly, the same %applies to the case of spinful particles.

\section*{Acknowledgments}
CS acknowledges financial support from the Swiss National Science Foundation (Grant P2EZP2 152190).

\section*{References}
%\bibliography{dualityBibliography}

\end{document}